\documentclass[epj-spec]{svjour}
\usepackage{graphicx}
\usepackage{amsmath}
\usepackage{amssymb}
\def\Nf{N_{\rm f}}

\def\mps{m_{\rm ps}}
\def\mpi{m_{\rm \pi}}
\def\mN{m_{\rm N}}

\newcommand{\Dlr}{\overset{\leftrightarrow}{D}}
\def\msbar{\overline{\rm MS}}
\begin{document}
%
%
\title{Hadronic structure from the lattice}
\author{%
Dirk Br\"{o}mmel\inst{1} \and
Meinulf G\"{o}ckeler\inst{2} \and
Philipp H\"{a}gler\inst{3} \and
Roger Horsley\inst{4} \and
Yoshifumi Nakamura\inst{5} \and
Munehisa Ohtani\inst{2} \and
Dirk Pleiter\inst{5}\fnmsep\thanks{\email{dirk.pleiter@desy.de}} \and
Paul~E.L.~Rakow\inst{6} \and
Andreas Sch\"{a}fer\inst{2} \and
Gerrit Schierholz\inst{5} \and
Wolfram Schroers\inst{7} \and
Hinnerk St\"uben\inst{8} \and
James M.~Zanotti\inst{4}
}
\institute{%
School of Physics and Astronomy, University of Southampton,
	Southampton SO17 1BJ, UK \and
Institut f\"ur Theoretische Physik, Universit\"at Regensburg,
	93040 Regensburg, Germany \and
Institut f\"ur Theoretische Physik T39, TU M\"unchen,
	85747~Garching, Germany \and
School of Physics, University of Edinburgh, Edinburgh EH9 3JZ, UK \and
Deutsches Elektronen-Synchrotron DESY and John von Neumann Institut f\"ur
	Computing NIC, 15738 Zeuthen, Germany \and
Theoretical Physics Division, Department of Mathematical Sciences,
	University of Liverpool, Liverpool L69 3BX, UK \and
Department of Physics, Center for Theoretical Sciences,
	National Taiwan University, Taipei 10617, Taiwan \and
Konrad-Zuse-Zentrum f\"ur Informationstechnik Berlin,
	14195 Berlin, Germany
}
\abstract{%
\vspace*{-3mm}
In recent years the investigation of hadron structure using lattice techniques
has attracted growing attention.  The computation of several important
quantities has become feasible. Furthermore, theoretical
developments as well as progress in algorithms and an increase in
computing resources have contributed to a significantly improved control of
systematic errors.  In this article we give an overview on the work
that has been carried out in the framework of the Hadron Physics I3
(I3HP) network ``Computational (lattice) hadron physics''. Here we will
not restrict ourselves to spin physics but focus on results for nucleon
spectrum and structure from the QCDSF collaboration.  For a broader
overview of developments in this field see, e.g., \cite{Hagler:2007hu}.
}
\maketitle
\vspace*{-3mm}

\section{Introduction}
\vspace*{-3mm}

The main goal of lattice hadron phenomenology is the computation of
generalized parton distributions (GPDs) of mesons and nucleons from
first principles. GPDs have become a theoretical framework which
enables us to study many fundamental aspects of the intrinsic hadron
structure. It allows us to confront experimental results and lattice
calculations since it includes as limiting cases hadron form factors
as well as polarized and unpolarized parton densities, i.e.~quantities which
are investigated by various experiments at CERN (COMPASS), DESY (H1,
Zeus, Hermes) and JLab. Combining information from both experiment and
lattice offers new opportunities to explore hadrons as extended objects.

On the lattice we are restricted to the calculation of moments of GPDs.
As an example we consider the unpolarized GPDs
$H^q(x,\xi,Q^2)$ and $E^q(x,\xi,Q^2)$. Their moments are related to
the generalised form factors (GFFs) $A_{n,k}^q(Q^2)$,
$B_{n,k}^q(Q^2)$ and $C_{n}^q(Q^2)$:
\vspace*{-3mm}
\begin{multline}
\int_{-1}^{+1} dx\,x^{n-1}
\left[
\begin{matrix}
H^q(x,\xi,Q^2) \\
E^q(x,\xi,Q^2) \\
\end{matrix}
\right]
\\ =
\sum_{i=0}^{[\frac{n-1}{2}]}
\left[
\begin{matrix}
A_{n,2i}^q(Q^2) \\
B_{n,2i}^q(Q^2) \\
\end{matrix}
\right]
(-2 \xi)^{2 i}
\pm {\rm Mod}(n+1,2) C_{n}^q(Q^2)\;(-2 \xi)^n\,.
\end{multline}
The GFFs can be calculated from a parametrisation of (nucleon) matrix elements.
As an illustrative example one may consider the GFFs $A^q_{1,0}$ and
$B^q_{1,0}$:
\begin{equation}
\label{eq:ma}
\left\langle p^{\prime},s^{\prime}\right|
 \bar{u} \gamma^{\mu} u
\left| p, s\right\rangle
=
\,\,\overline u(p',s ') \bigg\{ \gamma^{\mu}A_{1,0}^{(u)}(Q^2)
+i \sigma^{\mu \nu} \frac{q_\nu} {2 m_N} B_{1,0}^{(u)}(Q^2) \bigg\} u(p,s)\,,
\end{equation}
where $p$ ($s$) and $p'$ ($s'$) denote initial and final momenta
(spins), $q = p' - p$ the momentum transfer (with $Q^2=-q^2$).
Instead of calculating the matrix elements (i.e., the l.h.s.~of
Eq.~(\ref{eq:ma})) in Minkowski space-time, we calculate these within a
discretized Euclidean space-time by using appropriate combinations of
two- and three-point correlation functions. A typical combination is the
following ratio:
\begin{equation}
 R(t,\tau,p',p) =  
\frac{C_3(t,\tau,p',p)}{C_2(t,p')} \times
 \left[ \frac{C_2(\tau,p') C_2(t,p') C_2(t-\tau,p) }
{C_2(\tau,p) C_2(t,p) C_2(t-\tau,p')} 
\right]^{1/2}\,,
\end{equation}
where $C_2(t,p)$ is the unpolarized two-point function with a source at time $0$
and sink at time $t$, while the three-point function $C_3(t,\tau,p',p)$ has an
(local) operator $\cal{O}$ insertion at time $\tau$. The ratio
$R(t,\tau,p',p)$ is expected to become constant when keeping
sufficiently far away from source and sink, i.e.~for
$0 \ll \tau \ll t \lesssim \frac{1}{2} L_T$. For further details see,
e.g., \cite{gpd}.

\subsection{Lattice obstacles}

\begin{figure}[t]
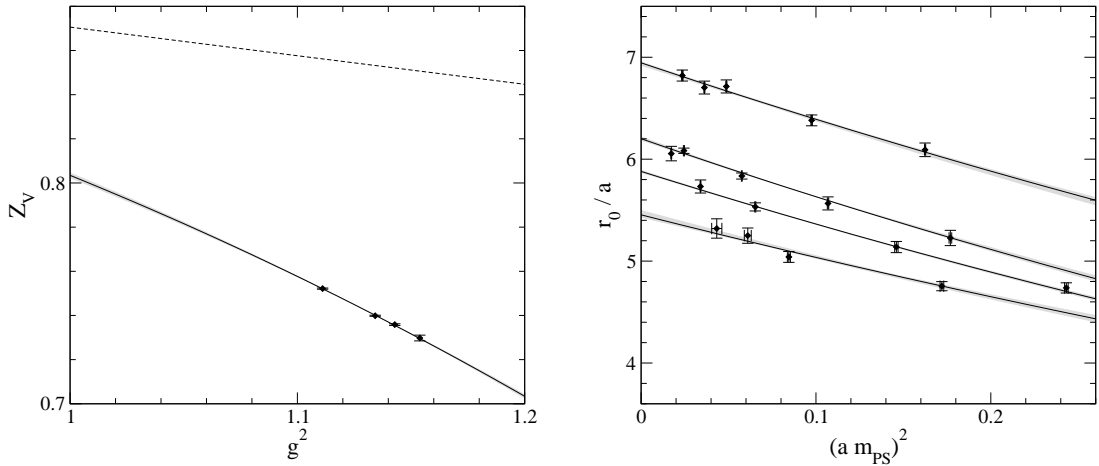

\includegraphics[scale=0.35]{figs/ZV.eps}\hfill
\includegraphics[scale=0.35]{figs/r0.eps}
\caption{The left plot shows a comparison of $Z_V$ obtained from 1-loop
perturbation theory (dashed line) and a non-perturbative determination
(solid line).
The right plot shows the values for $r_0(\beta,\mps)$
($\beta=5.20, 5.25, 5.29, 5.40$) together with an
extrapolation to the chiral limit.}
\label{fig:ZV}
\label{fig:r0}
\end{figure}

All lattice calculations are faced with the problem of keeping their
systematic errors under control. The main sources for such systematic
errors are the following:
\begin{itemize}
\item In almost all cases lattice results have to be extrapolated to the
  \begin{itemize}
  \item infinite volume limit: $V\rightarrow \infty$.
  \item continuum limit: $a\rightarrow 0$ where $a$ is the lattice spacing.
  \item region of light quark masses: $\mps\rightarrow m_{\pi}$.
  \end{itemize}
\item Renormalization and mixing of operators.
\item Conversion of the lattice results into physical units.
\end{itemize}

The results which we present in this article have been obtained from
simulations with $\Nf=2$ flavours of dynamical
$O(a)$-improved Wilson fermions (so-called Clover fermions).
Improvement of both the action and the operators is expected to
bring discretization errors down to $O(a^2)$. Our
simulations have been performed with lattice spacings in the range
of $0.11$ down to $0.07\,\mbox{fm}$. For most of our results we
find discretization effects to be small and often negligible compared to,
e.g., statistical errors.

For some of the observables considered here, finite volume effects turned
out to be large. In our simulations the lattices have a spatial extension
in the range $L_s = 1.4, \cdots, 2.6\,\mbox{fm}$.  Results from chiral
effective field theories may be used to estimate or even correct for finite
size effects.

Typically, the largest uncertainty originates from the extrapolation of
the results down to the region of physical light quark masses. Results
presented here have been obtained for pseudo-scalar meson masses in the
range of $300\,\mbox{MeV}$ up to $1\,\mbox{GeV}$. Due to a combination
of significantly improved algorithms (see \cite{Clark:2006wq} for a recent
review on algorithms for simulations with dynamical fermions) and the
availability of increased computer power it has become possible to start
exploring a quark mass region where one may hope results from chiral
perturbation theory to be applicable.

Another source of systematic errors stems from renormalization.
Perturbative renormalization techniques suffer from their complexity
when going beyond 1-loop and, more importantly, from their poor
convergence. It turned out to be crucial to employ non-perturbative
renormalization techniques. As an illustrative example we compare
in Fig.~\ref{fig:ZV} (left panel) the vector-current renormalization constant
$Z_V$ obtained from 1-loop perturbation theory and a non-perturbative
determination, where we imposed as renormalization condition the
equivalence of the local and the conserved vector current
(see \cite{Bakeyev:2003ff} for a detailed description of this method).

Let us finally consider the problem of converting lattice results into
physical units. It has become popular to use the Sommer parameter $r_0$,
which is defined by the expression $r_0^2\,F(r_0) = 1.65$.
The force $F(r)$ can be determined with
rather high statistical accuracy from the static quark potential. This
strategy however suffers from the problem that the phenomenological value is
not very well known. Typically, $r_0 \simeq 0.5\,\mbox{fm}$ is used.
Lattice results, e.g., for the nucleon mass (see \cite{dp-lat07}),
suggest a significantly smaller value. Here we will use
$r_0 = 0.467\,\mbox{fm}$ \cite{qcdsf-mn,Aubin:2004wf}.
Fig.~\ref{fig:r0} (right panel) shows our results for $r_0$.

\section{Masses}\label{sec:mass}

\begin{figure}[t]
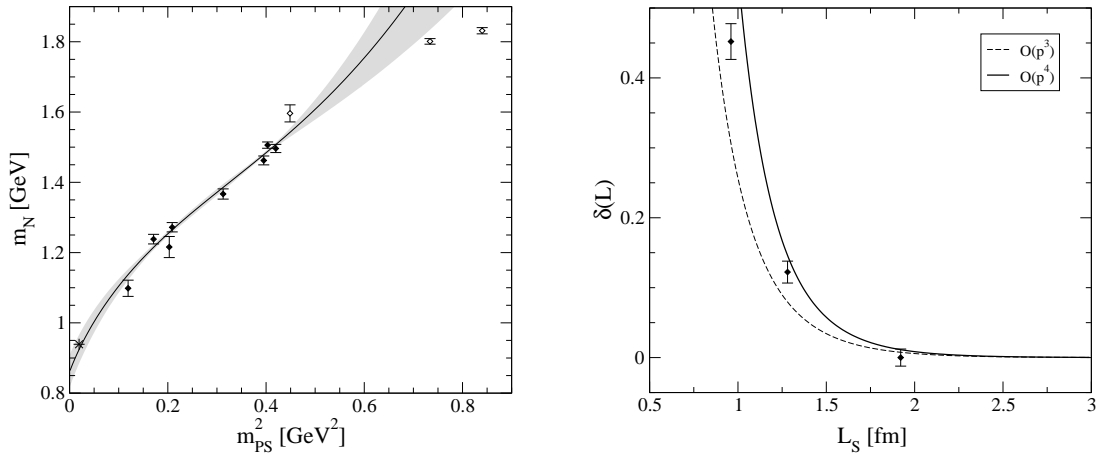

\includegraphics[scale=0.35]{figs/fitMnBchpt}\hfill
\includegraphics[scale=0.34]{figs/fse-b5p29kp13590}
\caption{\label{fig:mn}%
The left plot shows lattice results for $\mN$ together with
a fit to Eq.~(\protect\ref{eq:mn}). The star indicates the physical point.
In the right plot we compare the relative finite size effect
$\delta(L) = (m_N(L) - m_N (\infty)) / {m_N(\infty)}$ as measured
on the lattice and predicted by B$\chi$PT.
The diamonds show results from lattices with $V=12^3\times 32$,
$16^3\times 32$, $24^3\times 48$ (left to right).}
\end{figure}

As a first example for using chiral effective field theories in order to
guide extrapolation to physical quark masses we consider the mass of the
nucleon, $\mN$.  Calculations using relativistic baryon chiral perturbation
theory (B$\chi$PT) suggest 
a rather non-trivial quark mass dependence. The $p$-expansion at $O(p^4)$
in an infinite volume gives the following quark mass dependence
\cite{Procura:2003ig}:
%
\begin{multline}
\label{eq:mn}
 \mN(\mps)  =  {M_0} - 4 {c_1} \mps^2-
\frac{3 {g_{A,0}}^2}{32 \pi {F_0^2}} \mps^3 + \\
 \left[{e_1^r(\lambda)}-\frac{3}{64 \pi^2 {F_0}^2}
     \left( \frac{{g_{A,0}}^2}{M_0} -
     \frac{c_2}{2} \right) - \right.
\left.
   \frac{3}{32 \pi^2 {F_0}^2}
       \left( \frac{{g_{A,0}}^2}{M_0} - 8{c_1} +
       {c_2} + 4 {c_3} \right)
   \ln{\frac{\mps}{\lambda}} \right] \mps^4 \\
 + \;\;\frac{3 {g_{A,0}}^2}{256 \pi {F_0}^2 {M_0}^2}\mps^5 + O(\mps^6)\,.
\end{multline}

Even using all our lattice measurements at quark masses in the
the range $0 < \mps \lesssim 650\,\mbox{MeV}$ it is not possible to
sufficiently constrain all parameters in Eq.~(\ref{eq:mn}).
We therefore reduce the number of free fit parameters to
the nucleon mass in the chiral limit $M_0$, the not very well known
low-energy constant (LEC) $c_1$ and the counter-term
$e_1^r(\lambda)$ (we use $\lambda = 1\,\mbox{GeV}$). For all other parameters,
i.e.~the LECs $c_2$ and $c_3$, the pion decay constant $F_0$ and
the nucleon axial coupling $g_{A,0}$, we use phenomenological estimates.
Both lattice data and the resulting fit are shown in Fig.~\ref{fig:mn}
(left panel).
The lattice data seems to fall on a universal curve indicating
discretisation effects to be small, which we thus ignored.
We observe that $\mN(\mps=m_{\pi})$ is consistent with experiment.
Furthermore, we find $c_1 = -1.02(7)\,\mbox{GeV}^{-1}$, a value
which is consistent with other estimates~\cite{Bernard:2007zu}.

B$\chi$PT has also been used to calculate the finite size
effects \cite{qcdsf-mn}:
\begin{equation}
\label{eq:mnFse}
m_N(L_s) - m_N (\infty) = \Delta_a(L_s) +  \Delta_b(L_s) + O(p^5),
\end{equation}
where $\Delta_a$ represents the $O(p^3)$ result for the volume dependence
and $\Delta_b$ additional contributions at $O(p^4)$. All
coefficients in $\Delta_a(L_s)$ and $\Delta_b(L_s)$ are also present in
Eq.~(\ref{eq:mn}). It is thus possible to estimate the finite volume effects
obtained from Eq.~(\ref{eq:mnFse}) using the coefficients obtained from
a fit to Eq.~(\ref{eq:mn}). These estimates can then be used for
a comparison with
values of the nucleon mass computed on lattices that have a smaller
(physical) volume. Fig.~\ref{fig:mn} (right panel) shows such a comparison
at fixed $\mps \simeq 590\,\mbox{MeV}$.

\section{Nucleon axial coupling}\label{sec:gA}

\begin{figure}[t]
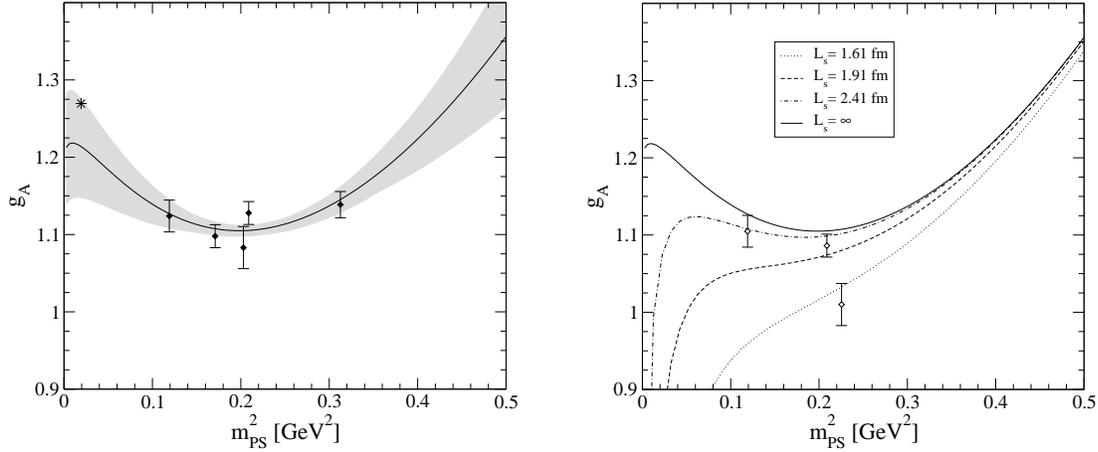

\includegraphics[scale=0.35]{figs/fitGaBchpt-beta-r0c}\hfill
\includegraphics[scale=0.34]{figs/gA-fse-plot}
\caption{\label{fig:gA}%
The left plot shows $g_A(\mps)$ in the infinite volume limit and the right
plot the data points before correcting the finite size effects.}
\end{figure}

Next, we consider the axial coupling constant $g_A = G_A(0)$, where
$G_A(Q^2)$ is the axial form factor of the nucleon. It is determined
from the renormalized axial vector current $A_{\mu}^R = Z_A\,(1+b_A\,a
m_q) A_{\mu}$, where $a m_q$ is the bare quark mass.  Here we only look
into the iso-vector case where contributions of so-called disconnected
terms cancel.  These are contributions of diagrams which only interact
with the hadron via the exchange of gluons and which are hard to compute
on the lattice.  While $Z_A$ is known non-perturbatively \cite{qcdsf-ga},
$b_A$ is only known perturbatively and is computed using tadpole improved
one-loop perturbation theory.

Like for the nucleon mass the quark mass dependence of the iso-vector
nucleon axial coupling has been calculated using chiral effective
field theory \cite{Hemmert:2003cb,qcdsf-ga}. The
calculations using the small scale expansion (SSE) have been performed
in the infinite volume limit as well as for a finite spatial cubic
box of length $L_s$.
We define $g_A(\mps, L_s) = g_A(\mps) + \Delta g_A(\mps, L_s)$,
where $\Delta g_A(\mps, L_s)$ denotes the finite size effects.
Given the large number of parameters in the resulting expressions we
again have to fix some of them using phenomenological input like the
pion decay constant, the $N\Delta$ mass splitting and the axial
$N\Delta$ coupling.

We find the finite size effects to be significantly larger compared to the
nucleon mass. These effects can not be ignored and we have to resort
to a fit which includes both quark mass dependence and finite size effects.
In Fig.~\ref{fig:gA} we show results for the quark mass dependence in the
infinite volume limit and finite size effects.

For results for $G_A(Q^2)$ at non-zero $Q^2$ and
for the pseudo-scalar form factor $G_P(Q^2)$ see \cite{Gockeler:2007hj}.
For an independent calculation of $g_A$ see \cite{Edwards:2005ym}.

\section{Electro-magnetic form factors}

JLab polarisation experiments \cite{JLabHallA,Punjabi:2005wq} have in
recent years led to a revived interest in the nucleon electro-magnetic
form factors. Measurements of the ratio of the proton electric to magnetic
form factors, $\mu^{\rm (p)}G_e^{\rm (p)}(Q^2)/G_m^{\rm (p)}(Q^2)$,
showed an unexpected decrease.  This means that the proton's electric
form factor falls off faster than the magnetic form factor.

The form factors are obtained from the standard decomposition of the
nucleon electromagnetic matrix elements
\begin{equation}
\langle p',s' | J^\mu | p,s \rangle =
\overline{u}(p',s')
\left[
  \gamma_\mu F_1^{(u)}(Q^2) + i\sigma^{\mu\nu}\frac{q_\nu}{2 m_N} F_2^{(u)}(Q^2)
\right]
u(p,s)\,.
\end{equation}
By calculating the matrix elements on the
l.h.s. and the nucleon mass we obtain the Dirac
form factor $F_1(Q^2)$ and the Pauli form factor $F_2(Q^2)$.

\begin{figure}[t]
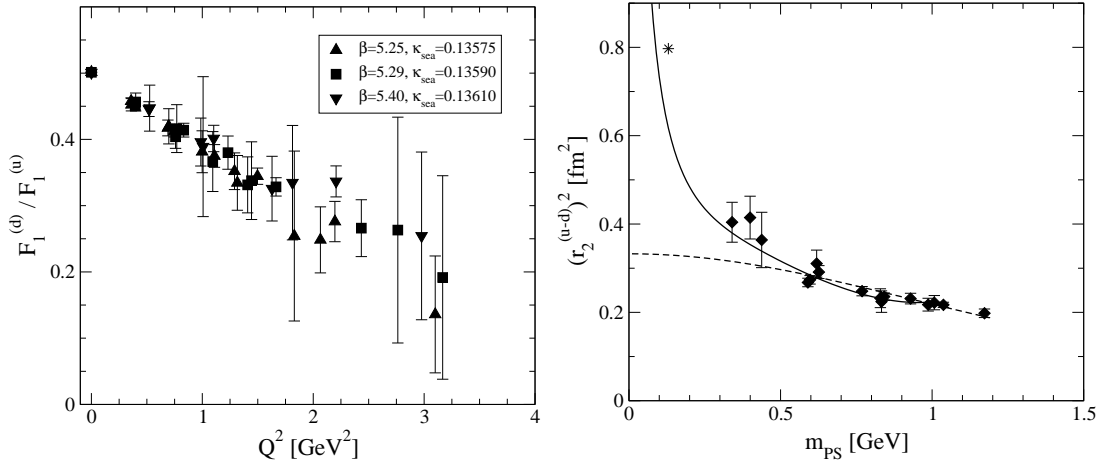

\includegraphics[scale=0.35]{figs/f1_f1_const_mpi}\hfill
\includegraphics[scale=0.35]{figs/r2v}
\caption{\label{fig:ff}%
The left plot shows the ratio $F_1^{(d)}/F_1^{(u)}$ as a function of $Q^2$
for different $\beta$ values but similar $\mps \simeq 400\,\mbox{MeV}$.
The right plot shows the extrapolation of our results for the Dirac
form factor radius $r_2^{(u-d)}$. The solid line shows a fit with the SSE
expression, the dashed line a fit to an ansatz linear in the quark mass.}
\end{figure}

Particularly interesting are the $Q^2$ scaling and flavour dependence of
the form factors. We find that the lattice data can be well parametrized
using a pole ansatz
\begin{equation}
F_i(Q^2) = \frac{A_i}{(1 + Q^2/M_i^2)^p}\,.
\end{equation}
(For a different ansatz see \cite{Gockeler:2007hj}.)
Naively, one would expect $p=2$ for $F_1$ and $p=3$ for $F_2$.
Our lattice data, however, has a flavour dependence which favours
$p=2$ for $F_1^{(u)}$ and $p=3$ otherwise.  In Fig.~\ref{fig:ff} (left panel)
we plot the ratio $F_1^{(d)}/F_1^{(u)}$ which one
naively would expect to be constant.
This result should be taken with some care since disconnected
contributions have not been calculated. These contributions only cancel in
the iso-vector case.  Our observation is however consistent with the flavour
dependence observed in fits to experimental data \cite{Diehl:2004cx}.

For a more quantitative comparison with experimental data we have to
extrapolate our results to physical quark masses. We again do this on the
basis of results from chiral effective field theories using the SSE
\cite{Hemmert:2002uh,Gockeler:2003ay}.
These calculations predict a strong quark mass dependence for the
iso-vector form factor radii and the iso-vector anomalous magnetic moment
in the small quark mass region.
In Fig.~\ref{fig:ff} (right panel) we compare our results for the iso-vector
Dirac radius $r_2^{(u-d)}$ and a fit to the SSE expression.
The lattice results are significantly smaller than the experimental
value. However, the SSE results indicate a strong quark mass dependence
for $\mps \lesssim 250\,\mbox{MeV}$, a region which is currently difficult
to access in lattice simulations.

\section{Moments of unpolarized structure functions}

As another quantity which can be calculated on the lattice we consider
the lowest moment of the unpolarized nucleon structure
function, $\langle x\rangle = A^q_{2,0}(0)$, where $A^q_{2,0}$ is the
first moment of $H^q(x,\xi,Q^2)$ at $\xi=0$.%

This moment is determined from the matrix element
\begin{equation}
\langle p,s |
\left[ \,{\cal O}^{\{{\mu}_1 {\mu}_2\}}\, - \mbox{Tr} \right]
| p,s \rangle =
2 A^q_{2,0} \left[p^{\mu_1} p^{\mu_2} - \mbox{Tr} \right]\,,
\end{equation}
where ${\cal O}^{\{\mu_1\mu_2\}} =
\overline{u}\,\gamma^{\mu_1} i \Dlr{}^{\mu_2} u$.
The renormalisation of this matrix element has been performed
non-perturbatively by means of the RI'-MOM method \cite{Z}. In the
perturbative conversion to $\msbar$ at a scale of 2 GeV we have used
$\Lambda^{\msbar} = 261(17)(26)\,\mbox{MeV}$ \cite{Gockeler:2005rv}.

\begin{figure}[t]
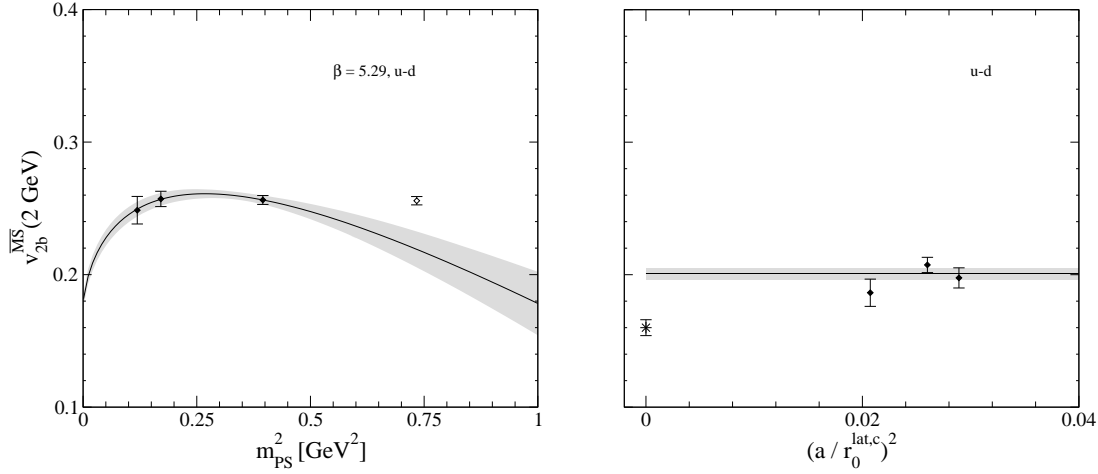

\begin{center}
\includegraphics[scale=0.35]{figs/x-umd-chiral}\hfill
\includegraphics[scale=0.35]{figs/x-umd-cont}
\end{center}
\caption{\label{fig:x-umd}%
The left plot shows our results for $\langle x\rangle^{(u-d)}(\mps)$ at
$\beta=5.29$ (the results for other values of $\beta$ are similar). The
solid line denotes a fit to a B$\chi$PT expression \cite{Dorati:2007bk}.
The right plot shows $\langle x\rangle^{(u-d)}(\mps=\mpi)$ as
a function of the lattice spacing together with a fit to a constant.
The star denotes the phenomenological value from CTEQ6.}
\end{figure}

In the range of quark masses which is currently accessible the results
in the iso-vector channel
$\langle x\rangle^{(u-d)}$ are significantly larger than the experimental
value. However, it has been suggested that this quantity may become
significantly smaller at very light quark masses \cite{Detmold:2001jb}.
This has been confirmed by recent calculations in the framework of baryon
chiral perturbation theory (B$\chi$PT) \cite{Dorati:2007bk}. These recent
calculations have been used to fit our results using the pion decay
constant, the nucleon mass and the nucleon axial coupling in the chiral
limit as phenomenological input. This leaves two free fit parameters,
which allows us to fit the data for different values of $\beta$ separately
restricting the fit range to $0 < \mps \lesssim 650\,\mbox{MeV}$.
In Fig.~\ref{fig:x-umd} we plot the fit to the $\beta=5.29$ data
showing little evidence for $\langle x\rangle^{(u-d)}(\mps)$ becoming
smaller for $\mps \rightarrow \mpi$. In our results we find no
indication for large discretisation effects (see right panel of
Fig.~\ref{fig:x-umd}). However, for the results at light
quark masses finite size effects cannot be excluded.

\begin{figure}[t]
\begin{center}
\includegraphics[scale=0.26,angle=-90]{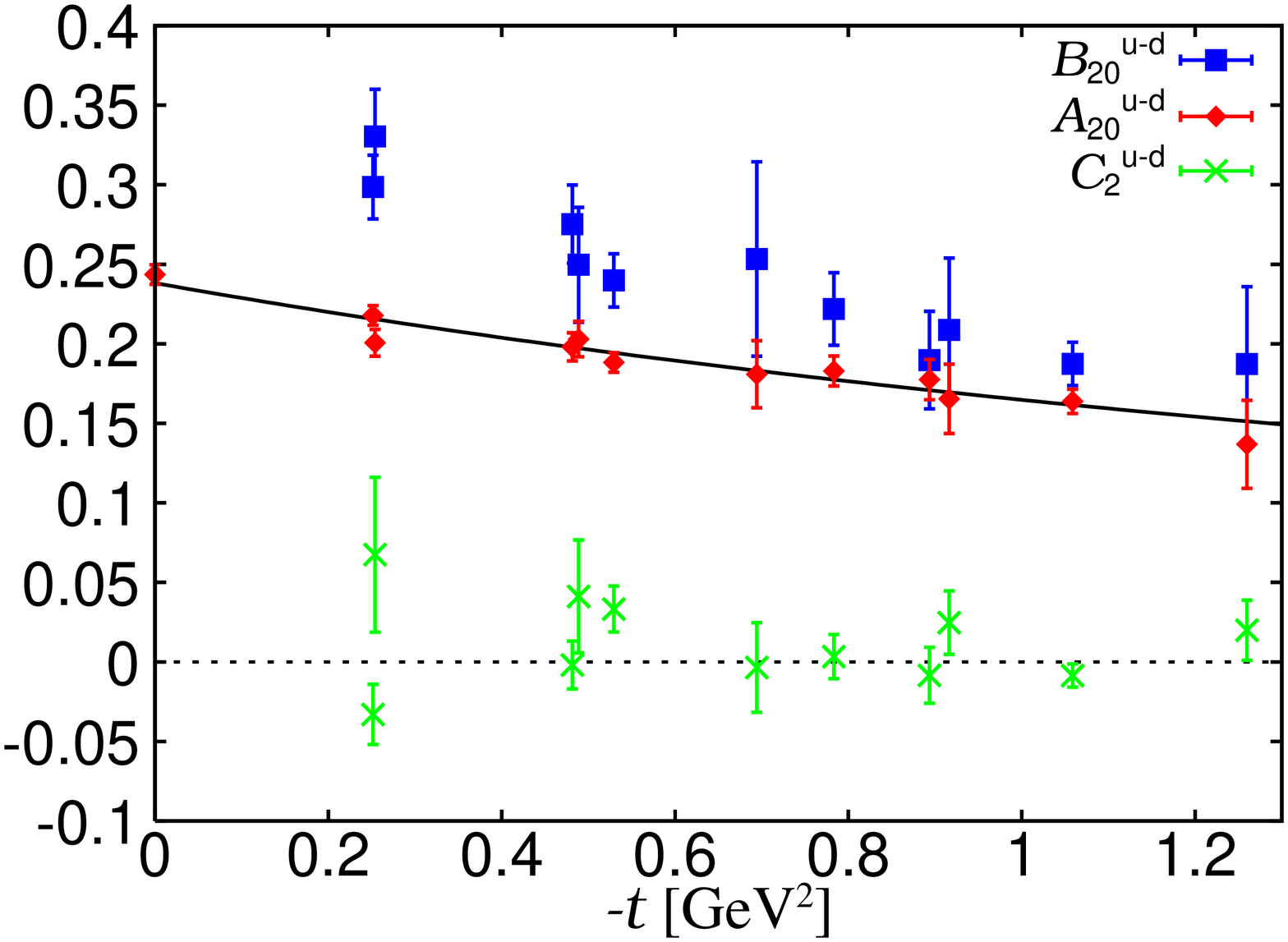}\hfill
\includegraphics[scale=0.26,angle=-90]{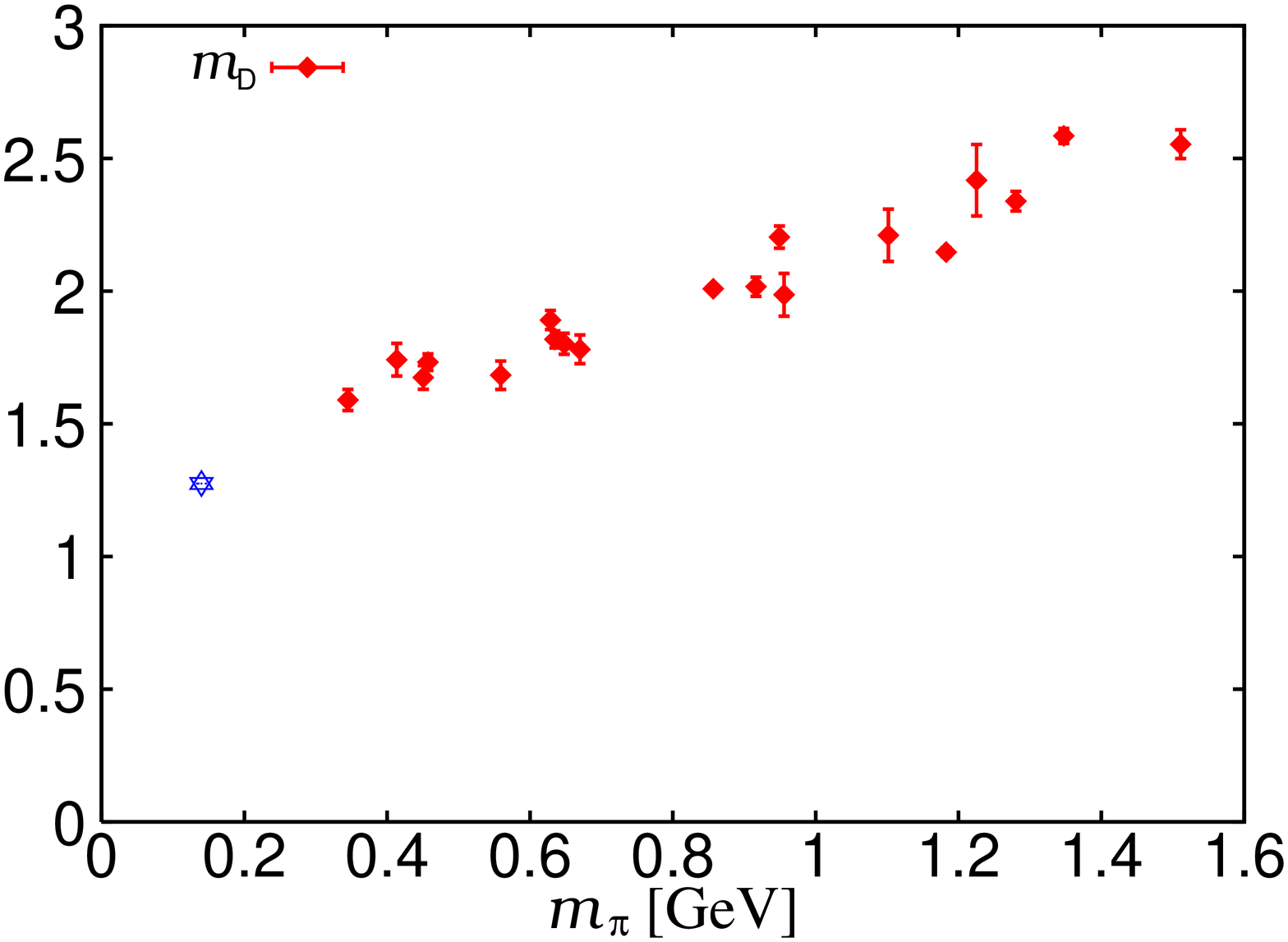}
\end{center}
\caption{\label{fig:gff}%
The left plot shows the
generalised form factors $A^q_{2,0}$, $B^q_{2,0}$ and $C^q_{2}$
in the iso-vector channel at $\kappa=0.13632$ and $\beta=5.29$
with $\mps \simeq 350\,\mbox{MeV}$.
In the right plot the dipole masses of $A^q_{2,0}$ are shown as a
function of the pseudo-scalar meson mass.
The star represents the experimental value of the tensor meson mass $f_2$.
}
\end{figure}

In Fig.~\ref{fig:gff} (left) the $Q^2$ dependence of the generalized form
factor $A^q_{2,0}$ is shown together with the other unpolarized GFFs
$B^q_{2,0}$ and $C^q_{2}$. The lattice results for $A^q_{2,0}$ can be well
described by a dipol expression 
$A^q_{2,0}(Q^2) = A^q_{2,0}(0) / (1 + Q^2/M_D^2)^2$. The quark mass
dependence of the dipole mass is also shown in Fig.~\ref{fig:gff} (right plot).
For further details see \cite{Brommel:2007sb}.

\section{Other results}

Several other results have been published in recent years, which we will
summarize here only very briefly:
\begin{itemize}
\item A calculation of the lowest two moments of transverse spin
densities of quarks in the nucleon has revealed strongly distorted
densities of transversely polarized quarks in the nucleon
\cite{Gockeler:2006zu}.

\item The electro-magnetic form factor $F_{\pi}$ of the pion calculated on
the lattice has been found to be in good agreement with experimental
results \cite{Brommel:2006ww}. The charge radius squared was obtained as
$\langle r^2\rangle = 0.441(19)\,\mbox{fm}^2$.

\item Lattice calculations of the first two moments of the quark tensor
GPD $E^{\pi}_T$ indicate -- like in the case of the nucleon -- a
strongly distorted spatial distribution of the quarks
if they are transversely polarized \cite{Brommel:2007xd}.

\end{itemize}

\section{Conclusions and outlook}

In this talk we have presented a small selection of results to demonstrate
where lattice calculations can contribute to the investigation of
hadron structure. While these results may improve our qualitative
understanding, for quantitatively precise results further efforts are
needed and keeping systematic errors under control remains a major challenge.
In particular, the necessary extrapolations to the
infinite volume, vanishing lattice spacing and to the physical quark
masses remain a major source of uncertainties. While results from chiral
effective field theories help to guide these extrapolations, eventually
simulations using large lattices and very light quark masses are needed.
We conclude with the good news that such simulations are now becoming feasible.

\section*{Acknowledgements}

The numerical calculations have been performed on the Hitachi SR8000 at
LRZ (Munich), the Cray T3E at EPCC (Edinburgh) the
APE{\it 1000} and apeNEXT at NIC/DESY (Zeuthen), the BlueGene/L at NIC/FZJ
(J\"ulich) and EPCC (Edinburgh). Some of the configurations at the small
pion mass have been generated on the BlueGene/L at KEK by the Kanazawa
group as part of the DIK research programme.  This work was supported in
part by the DFG, by the EU Integrated Infrastructure Initiative Hadron
Physics (I3HP) under contract number RII3-CT-2004-506078,
and by the National Science Council of Taiwan under the grant numbers
NSC96-2112-M002-020-MY3 and NSC96-2811-M002-026.


\end{document}